\documentclass[singlespacing]{elsart}

\usepackage{graphicx}

\usepackage{amssymb}
\journal{Computational and Theoretical Chemistry}
\begin{document}

\begin{frontmatter}

\title{Analytic treatment of IR-spectroscopy data for double well potential}

\author{ A.E. Sitnitsky},
\ead{sitnitsky@kibb.knc.ru}

\address{Kazan Institute of Biochemistry and Biophysics, FRC Kazan Scientific Center, Russian Academy of Sciences, P.O.B. 30,
420111, Russian Federation. Tel. 8-843-231-90-37. e-mail: sitnitsky@kibb.knc.ru }

\begin{abstract}
A theoretical scheme for the analysis of experimental data on IR spectroscopy for a quantum particle in a double well potential (DWP) is suggested. The analysis is based on the trigonometric DWP for which the exact analytic solution of the Schr\"odinger equation is available. The corresponding energy levels along with their wave functions are expressed via special functions implemented in {\sl {Mathematica}} (spheroidal function and its spectrum of eigenvalues). As a result trigonometric DWP makes the calculation of the energy levels an extremely easy procedure. It contains three parameters allowing one to model the most important characteristics of DWP (barrier height and the distance between the minima of the potential) along with the required asymmetry. Our approach provides an accurate calculation of the energy spectrum for hydrogen bonds in chromous acid (CrOOH) and potassium dihydrogen phosphate (${\rm{KH_2PO_4}}$) along with their polarizability in agreement with available experimental data.
\end{abstract}

\end{frontmatter}

\section{Introduction}
The one-dimensional time-independent Schr\"odinger equation (SE) for a quantum particle moving in some double-well potential (DWP) is ubiquitous in physics and chemistry (see \cite{Raz80}, \cite{Jel12}, \cite{Xie12}, \cite{Dow13}, \cite{Che13}, \cite{Har14}, \cite{Sit17}, \cite{Sit171}, \cite{Sit18}, \cite{Don18}, \cite{Don181}, \cite{Don182}, \cite{Don183}, \cite{Ish16}, \cite{Kom76}, \cite{Agb11} and refs. therein). On the one hand numerical methods for the solution of SE (see \cite{Lev14}, \cite{Her15} and refs. therein) provide accurate description for both one-dimensional and multidimensional problems. On the other hand the exact analytic solutions \cite{Xie12}, \cite{Dow13}, \cite{Che13}, \cite{Har14}, \cite{Sit17}, \cite{Sit171}, \cite{Sit18}, \cite{Don18}, \cite{Don181}, \cite{Don182}, \cite{Don183}, \cite{Ish16}, \cite{Agb11} remain an important tool for understanding the reality. Such solutions are expressed via confluent Heun's function (CHF) \cite{Xie12}, \cite{Dow13}, \cite{Che13}, \cite{Har14}, \cite{Sit17}, \cite{Sit171}, \cite{Sit18}, \cite{Don18}, \cite{Don181}, \cite{Don182}, \cite{Ish16} (implemented in {\sl {Maple}}), spheroidal function (SF) \cite{Sit17}, \cite{Sit171}, \cite{Sit18} or derived via the functional Bethe ansatz \cite{Agb11}. Unfortunately these solutions can be obtained only for some particular DWPs and lack the universality of numerical methods. Trigonometric DWP
\[
U(x)=\left(m^2-\frac{1}{4}\right)\ \tan^2 x -p^2\ \sin^2 x-a \sin x
\]
where $-\pi/2\leq x \leq \pi/2$ is a particular type of DWPs expressed via trigonometric functions for which SE can be exactly solved via SF (implemented in {\sl {Mathematica}}) or CHF \cite{Sit17}, \cite{Sit171}, \cite{Sit18}. In the symmetric form $a=0$ (see examples in Fig.1, Fig.2 and Fig.5) it contains two parameters $\{m,p\}$ allowing one to model the most important characteristics of DWP (barrier height and the distance between the minima of the potential). In the asymmetric form it contains the third (asymmetry $a\not= 0$) parameter. The merit of trigonometric DWP is in the fact that SF \cite{Kom76} is implemented in {\sl {Mathematica}} along with its spectrum of eigenvalues. As a result the calculation of the energy levels becomes an easy procedure. It is obvious that not all DWPs (in particular those complicated curves of quantum chemical calculations) can be covered by the trigonometric form. On the other hand there are cases in which the IR spectroscopy data can be accurately described by this DWP. Thus it is an additional and complementary tool in the toolkit of the researchers dealing with IR spectroscopy. It enables one to make quick calculations of the energy levels and the wave functions of the system under investigation from the IR spectroscopy data. Also with the rational time consumption it allows one to obtain the characteristics of the system calculated on the basis of the above information, e.g., the behavior of its polarizability on temperature (see below). In the present article we apply SE with trigonometric DWP to the analytic description of the IR-spectroscopy data for symmetric hydrogen bonds (HB) in chromous acid (CrOOH) and potassium dihydrogen phosphate (${\rm{KH_2PO_4}}$).

SE with DWP is often used for the description of the proton position for HB \cite{Law801}, \cite{Law81}, \cite{Law87}, \cite{Wie70}, \cite{Jan72}, \cite{Kry91}, \cite{Eck87}, \cite{Zun00}. The aim of the present article is to show that trigonometric DWP  provides the expressions for the energy levels and their corresponding wave functions for HB in CrOOH and ${\rm{KH_2PO_4}}$ along with their polarizability. The tunneling process for a proton is known to yield a dominant contribution {\bf{to}} the extremely high polarizability of HB \cite{Wie70}, \cite{Jan72}, \cite{Kry91},  \cite{Eck87}, \cite{Zun00}. This problem was thoroughly investigated by Zundel and co-authors (see \cite{Wie70}, \cite{Eck87}, \cite{Zun00} and refs. therein). In \cite{Kry91} the ground-state splitting was calculated by the instanton approach to double Morse potential. For low-barrier HB in CrOOH and ${\rm{KH_2PO_4}}$ the ground state doublet is close to the barrier top and the instanton method is known to yield very inaccurate results \cite{Sit18} in this case. Our approach provides a more stringent and adequate tool for calculating the ground state splitting. However it should be stressed that the intricate phenomenon of the polarizability of realistic HB is known to be complicated by many other factors arising from their interaction with environment \cite{Kry91},  \cite{Zun00}. Besides we are not aware of the experimental data on the polarizability of HB in CrOOH or ${\rm{KH_2PO_4}}$. Thus we can not compare our model calculations with the experiment. Nevertheless the calculation of the polarizability for HB in CrOOH and ${\rm{KH_2PO_4}}$ at room temperature yields the values  very close to that calculated from the available experimental data for HB in BrH $\cdot \cdot \cdot$ N \cite{Eck87}.$\ $In our opinion such coincidence can not be fortuitous and is a serious argument in favor of the relationship of trigonometric DWP to the reality.

\section{Energy levels, wave functions and polarizability for quantum particle in trigonometric double-well potential}
Polarizability $\alpha \left(T\right)$ as a function of temperature $T$ is a reaction of the system under consideration to a constant external electric field making DWP intrinsically asymmetric \cite{Eck87}. Thus a self-consistent approach should include an asymmetry term in the Hamiltonian $-\mu(X) F$ where $F$ is the electric field strength and $\mu(X)$ is the dipole moment. The latter is usually assumed to be $\mu(X)=e X$ in the linear approximation. In Appendix 1 we show how to transform the dimensional values into their dimensionless counterparts for trigonometric DWP. We denote the asymmetry parameter
\begin{equation}
\label{eq1} a=\frac{16ML^3eF}{\hbar^2 \pi^3}
\end{equation}
Here $e$ is the quantum particle charge, $M$ is the mass of the quantum particle, $\hbar$ is the reduced Planck constant and $L$ is the dimensional boundary of the interval for the spatial variable $X$ (see Appendix 1).
Then the dimensionless interaction energy is $-ax$. However the linear approximation can be valid within the interval of a sufficiently small x only. Common knowledge pervading textbooks on physical chemistry ascertains that the realistic behavior of the dipole moment must deflect to more slow growth than the linear one (see, e.g., Fig. 10.54 in \cite{Atk09}). To go beyond the framework of the linear approximation and to model such deflection we replace the linear term by the trigonometric one
\begin{equation}
\label{eq2} -ax\longrightarrow -a\sin x
\end{equation}
at $-\pi/2 \leq x \leq \pi/2$. The latter approximation is certainly better than the linear one, it is natural for our trigonometric DWP and finally it enables us to proceed with analytic treatment of the problem.

Then SE with trigonometric DWP in dimensionless units (see Appendix 1 for details) takes the form \cite{Sit171}, \cite{Sit18}
\begin{equation}
\label{eq3} \psi''_{xx} (x)+\left[\epsilon-\left(m^2-\frac{1}{4}\right)\ \tan^2 x +p^2\ \sin^2 x+a \sin x\right]\psi (x)=0
\end{equation}
Here $-\pi/2\leq x \leq \pi/2$, $m$ (an integer number) is the barrier width parameter and $p$ is the barrier height parameter.
The solution of (\ref{eq3}) is discussed in details in \cite{Sit18}
\begin{equation}
\label{eq4} \psi_q (x)=\cos^{1/2} x\ \bar\Xi_{mq}\left(p, a;\sin x\right)
\end{equation}
Here $q=0,1,2,...$ and $\bar\Xi_{mq}\left(p, a;\sin x\right)$ is angular generalized (Coulomb) SF \cite{Kom76}. The energy levels are determined by the relationship
\begin{equation}
\label{eq5} \epsilon_q=\lambda_{mq}\left(p, a\right)+\frac{1}{2}-m^2-p^2
\end{equation}
Here $\lambda_{mq}\left(p, a\right)$ is the spectrum of eigenvalues for the function $\bar\Xi_{mq}\left(p, a;s\right)$.
\begin{figure}
\begin{center}
\includegraphics* [width=\textwidth] {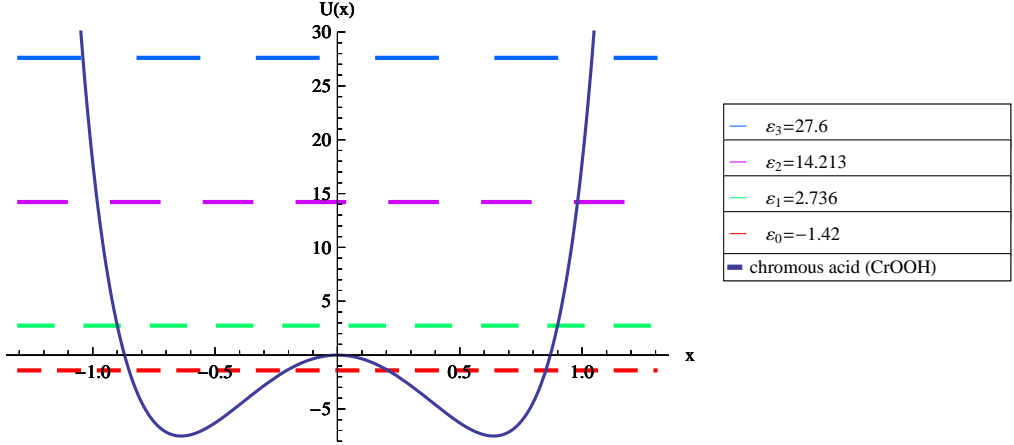}
\end{center}
\caption{The symmetric trigonometric double-well potential at the values of the parameters $m=5$, $p=7.716$. The parameters are chosen to describe the potential and the energy levels for the hydrogen bond in chromous acid (CrOOH)
(experimental data are taken from \cite{Law87}). The energy levels $\epsilon_0=-1.42$, $\epsilon_1=2.736$, $\epsilon_2=14.213$, $\epsilon_3=27.6$ are respectively depicted by the dashes of increasing length. The splitting of the ground state $\epsilon_1-\epsilon_0=4.156$ corresponds to $546\ {\rm cm^{-1}}$ in dimensional units.} \label{Fig.1}
\end{figure}
\begin{figure}
\begin{center}
\includegraphics* [width=\textwidth] {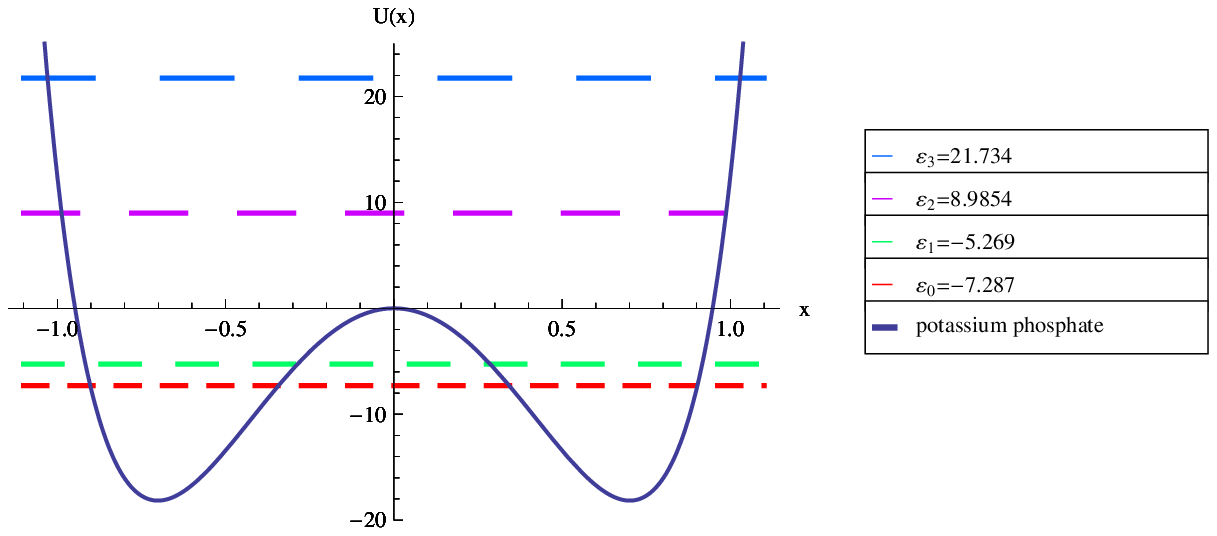}
\end{center}
\caption{The symmetric trigonometric double-well potential at the values of the parameters $m=6$, $p=10.24$. The parameters are chosen to describe the potential and the energy levels for the hydrogen bond in potassium dihydrogen phosphate (${\rm{KH_2PO_4}}$)
(experimental data are taken from \cite{Law801}, \cite{Law81}). The energy levels $\epsilon_0=-7.287$, $\epsilon_1=-5.269$, $\epsilon_2=8.9854$, $\epsilon_3=21.734$ are respectively depicted by the dashes of increasing length. The splitting of the ground state $\epsilon_1-\epsilon_0=2.018$ corresponds to $320\ {\rm cm^{-1}}$ in dimensional units.} \label{Fig.2}
\end{figure}

With the solution (\ref{eq4}) further calculations are straightforward along the general formulas for the dipole moment and the definition of the polarizability \cite{Eck87}. The average dipole moment in the $n$-th state with taking into account the replacement (\ref{eq2}) and with the help of (1.24b) from the part 2 in \cite{Kom76} is
\begin{equation}
\label{eq6} \mu_{nn}=-\int_{-\pi/2}^{\pi/2}dx\ \psi_n (x)\ \sin x\ \psi_n (x)=-\int_{-1}^{1}d\eta\ \eta\left[\bar\Xi_{mn}\left(p, a;\eta\right)\right]^2=\frac{\partial \lambda_{mn}\left(p, a\right)}{\partial a}
\end{equation}
We denote the dimensionless inverse temperature $\beta$ and the dimensionless polarizability $\Theta\left(\beta\right)$
\begin{equation}
\label{eq7}
\beta=\frac{\hbar^2\pi^2}{16ML^2k_BT}\ \ \ \ \ \ \ \ \ \ \ \ \ \ \ \ \ \ \ \ \ \ \ \ \Theta\left(\beta \right)=\frac{\hbar^2\pi^4}{64e^2L^4M}\alpha \left(T\right)
\end{equation}
Here $k_B$ is the Boltzmann constant and $T$ is the temperature.
The mean dipole moment as a function of the dimensionless inverse temperature is
\begin{equation}
\label{eq8} \mu\left(\beta, a\right)=\left[\sum_{n=0}^{\infty}\ \exp\left(-\beta \epsilon_n\right)\right]^{-1}\sum_{n=0}^{\infty}\ \frac{\partial \lambda_{mn}\left(p, a\right)}{\partial a}\exp\left(-\beta \epsilon_n\right)
\end{equation}
Finally the field zero polarizability is
\begin{equation}
\label{eq9}
\Theta\left(\beta\right)=\frac{\partial \mu\left(\beta, a\right)}{\partial a}\left| {\begin{array}{l}
  \\
{a=0}\\
 \end{array}}\right. \
\end{equation}
The stringent procedure for the calculation of $\lambda_{mn}\left(p, a\right)$ is extremely cumbersome \cite{Liu92}. However
the requirement to carry out the calculations in the limit $a \rightarrow 0$ enables us to simplify them considerably. For an approximate estimate we use the expression from \cite{Pon76}, \cite{Liu92} that is valid for $a \rightarrow 0$ at low values of $m$ and $p$. As a result we obtain
\begin{equation}
\label{eq10}\lambda_{mn}\left(p, a\right)\approx A_n+B_n \frac{a^2}{2}
\end{equation}
Here
\begin{equation}
\label{eq11} A_n=\lambda_{mn}\left(p, 0\right)\equiv \lambda_{m(n+m)}\left(p\right)
\end{equation}
where $\lambda_{m(q+m)}\left(p\right)\equiv SpheroidalEigenvalue[q+m,m,ip]$ is the spectrum of eigenvalues for SF written in  the designations of \cite{Kom76} and in {\sl {Mathematica}} notation. For $B_n$ we have from the above mentioned expression \cite{Pon76}, \cite{Liu92}
\begin{equation}
\label{eq12} B_n \approx \frac{1}{2(n+m)+1}\left[\frac{m^2-(n+m)^2}{(n+m)\left(2(n+m)-1\right)}+\frac{(n+m+1)^2-m^2}{(n+m+1)\left(2(n+m)+3\right)}\right]
\end{equation}
With the help of (\ref{eq10}) eq. (\ref{eq8}) takes the form
\begin{equation}
\label{eq13} \mu\left(\beta, a\right)\approx \left\{\sum_{n=0}^{\infty}\ \exp\left[-\beta \left(A_n+\frac{a^2B_n }{2}\right)\right]\right\}^{-1}a\sum_{n=0}^{\infty}\ B_n \exp\left[-\beta \left(A_n+\frac{a^2B_n }{2}\right)\right]
\end{equation}
The latter expression makes the polarizability (\ref{eq9}) to be calculable. The results obtained with the help of {\sl {Mathematica}}  \cite{Wol13} are presented in Fig.3$\ $  and Fig.4.
\begin{figure}
\begin{center}
\includegraphics* [width=\textwidth] {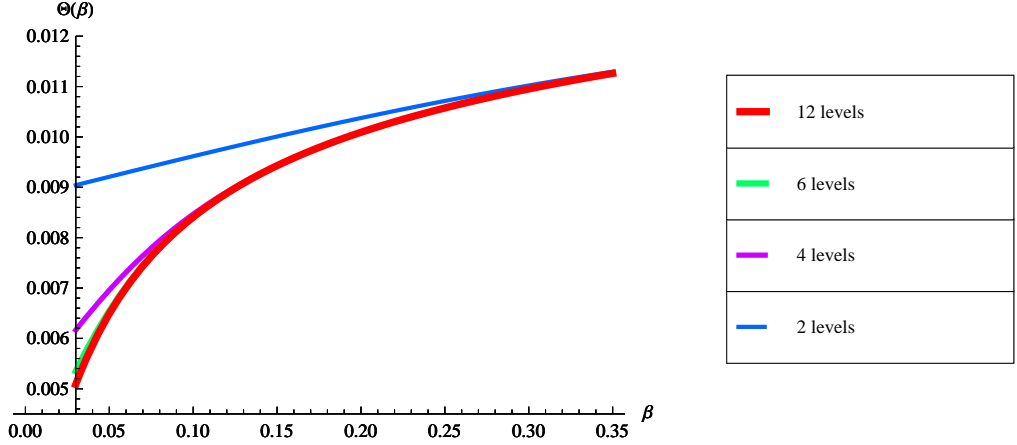}
\end{center}
\caption{The dependence of the dimensionless polarizability $\Theta\left(\beta\right)$ (\ref{eq9}) of the quantum particle in the trigonometric double-well potential on the dimensionless inverse temperature (\ref{eq7}) at the values of the barrier height parameter $p=7.716$ and the barrier width parameter $m=5$ corresponding to the hydrogen bond in chromous acid (CrOOH). The results for the increasing number of doublets (pairs of energy levels) taken into account in (\ref{eq13}) are depicted by the lines of increasing thickness respectively.} \label{Fig.3}
\end{figure}
\begin{figure}
\begin{center}
\includegraphics* [width=\textwidth] {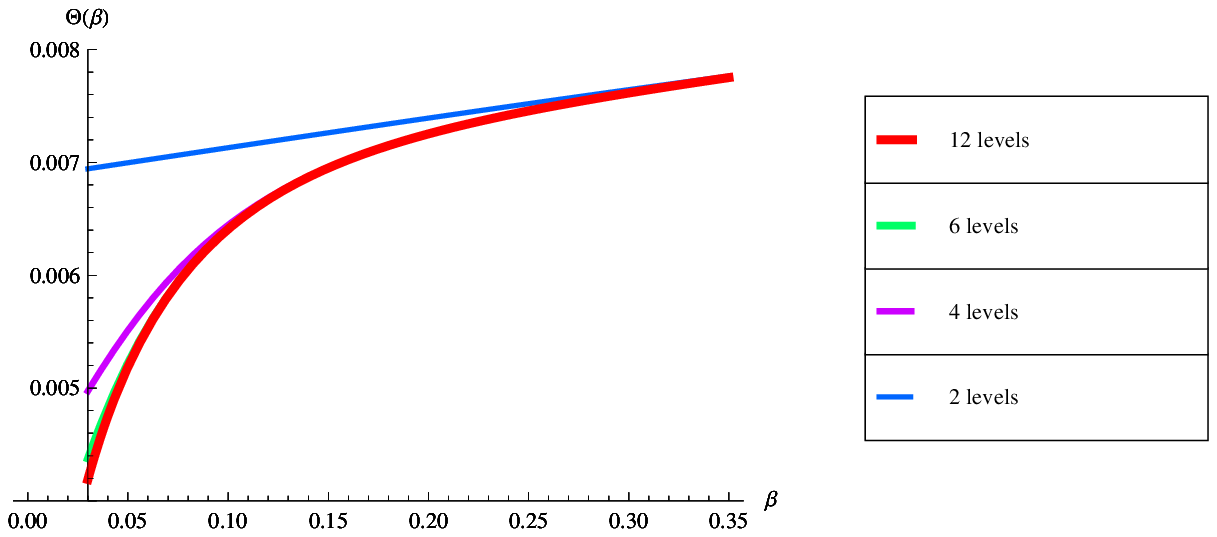}
\end{center}
\caption{The dependence of the dimensionless polarizability $\Theta\left(\beta\right)$ (\ref{eq9}) of the quantum particle in the trigonometric double-well potential on the dimensionless inverse temperature (\ref{eq7}) at the values of the barrier height parameter $p=10.24$ and the barrier width parameter $m=6$ corresponding to the hydrogen bond in potassium dihydrogen phosphate (${\rm{KH_2PO_4}}$). The results for the increasing number of doublets (pairs of energy levels) taken into account in (\ref{eq13}) are depicted by the lines of increasing thickness respectively.} \label{Fig.4}
\end{figure}

\section{Results and discussion}
In Fig.1 the energy levels for HB in chromous acid (CrOOH) (experimental data are taken from \cite{Law87}) are presented. In Fig.2 the energy levels for HB in potassium dihydrogen phosphate (${\rm{KH_2PO_4}}$) (experimental data are taken from \cite{Law801}, \cite{Law81}) are presented. For HB in CrOOH \cite{Law87} (${\rm{KH_2PO_4}}$ \cite{Law801}, \cite{Law81}) the ground-state splitting is $\Delta E=E_1-E_0\approx 546\ {\rm cm^{-1}}$ ($320\ {\rm cm^{-1}}$) while the frequencies for the higher levels are $E_2-E_1\approx 1508\ {\rm cm^{-1}}$ ($2260\ {\rm cm^{-1}}$), $E_3-E_0\approx 3802\ {\rm cm^{-1}}$ ($4600\ {\rm cm^{-1}}$) and $E_2-E_0\approx 2053\ {\rm cm^{-1}}$ ($2580\ {\rm cm^{-1}}$). The ratios
\[
\frac{E_2-E_0}{E_2-E_1}=\frac{\epsilon_2-\epsilon_0}{\epsilon_2-\epsilon_1}=1.36\ (1.14);\ \ \ \ \ \ \frac{E_2-E_1}{E_1-E_0}=\frac{\epsilon_2-\epsilon_1}{\epsilon_1-\epsilon_0}\approx 2.76\ (7.06);
\]
\[
\frac{E_3-E_0}{E_2-E_0}=\frac{\epsilon_3-\epsilon_0}{\epsilon_2-\epsilon_0}\approx 1.85\ (1.78)
\]
are obtained for our dimensionless energy levels $\epsilon_q$ if we take the values for the parameters of DWP $m=5\ (6)$ and $p=7.716\ (10.24)$.

Fig.3 and Fig.4 show the behavior of the polarizability with the decrease of temperature (increase of $\beta$) at the value of the parameters corresponding to HB in CrOOH and ${\rm{KH_2PO_4}}$ respectively. In Fig.3. and Fig.4 the number of doublets (pairs of energy levels) taken into account in (\ref{eq13}) means the quantity "$N$" used for the truncation of the infinite summation in (\ref{eq13}). For instance "6 levels" in Fig.3 and Fig.4 means that the series is truncated by $N=5$ (n=0,...,5) while "12 levels" means that the series is truncated by $N=11$ (n=0,...,11). The curves in Fig.3 and Fig.4 practically are not altered at increasing the truncation number over $N=5$ that proves the convergence of the series in (\ref{eq13}).

One can conclude that the suggested trigonometric double-well potential for the Schr\"odinger equation provides accurate analytic calculation of the energy levels,  the corresponding wave functions and the polarizability of hydrogen bonds in CrOOH and ${\rm{KH_2PO_4}}$ via special functions implemented in {\sl {Mathematica}}.

\section{Appendix 1}
In dimensional units the one-dimensional SE for a quantum particle with the reduced mass $M$ has the form
\begin{equation}
\label{eq14} \frac{d^2 \psi (X)}{dX^2}+\frac{2M}{\hbar^2}\left[E-V(X)\right]\psi (X)=0
\end{equation}
where $-L \leq X \leq L$ and $V(X)$ is a DWP. The latter is assumed to be infinite at the boundaries of the finite interval for the spatial variable $X=\pm L$.
The dimensionless values for the distance $x$, the potential $U(x)$ and the energy $\epsilon$ are introduced as follows
\begin{equation}
\label{eq15} x=\frac{\pi X}{2L}\ \ \ \ \ \ \ \ \ \ \ \ \ \ \ \ \ U(x)=\frac{8ML^2}{\hbar^2 \pi^2}V(X)\ \ \ \ \ \ \ \ \ \ \ \ \ \epsilon=\frac{8ML^2E}{\hbar^2 \pi^2}
\end{equation}
where $-\pi/2\leq x \leq \pi/2$. As a result we obtain dimensionless SE (\ref{eq3}) with trigonometric DWP.

For HB in both chromous acid \cite{Law87} and potassium dihydrogen phosphate \cite{Law801}, \cite{Law81} the distance between the oxygen atoms is $2L\approx2.49\ \AA$ so that for the boundary of trigonometric DWP we have $L\approx1.245\ \AA$. The mass of a proton is $M=1\ {\rm amu}$. From these values we estimate the inverse dimensionless temperature (\ref{eq7}) to be $\beta \approx 0.0583$ at $T=300\ K$ and  $\beta \approx 0.175$ at $T=100\ K$. The calculated value for the dimensionless polarizability in chromous acid at room temperature is $\Theta\left(0.0583\right)\approx 0.0068$ (see Fig.3). The calculated value for the dimensionless polarizability in potassium dihydrogen phosphate at room temperature is $\Theta\left(0.0583\right)\approx 0.0054$ (see Fig.4). These values are close to those calculated from the known experimental values: $\Theta\left(0.0583\right)\approx0.00676$ obtained from $\alpha \left(300\ K\right)=40 \cdot 10^{-24}\ cm^3$ for HB in BrH $\cdot \cdot \cdot$ N \cite{Eck87}.

\section{Appendix 2}
Another case of making use the trigonometric DWP takes place when the results of non-empirical quantum chemical (ab initio) potential energy surface calculations are available instead of the IR spectroscopy frequencies. To exemplify this case we apply the present approach to the Zundel ion ${\rm{H_5O_2^{+}}}$ (oxonium hydrate) basing on the results of \cite{Yu16}.
\begin{figure}
\begin{center}
\includegraphics* [width=\textwidth] {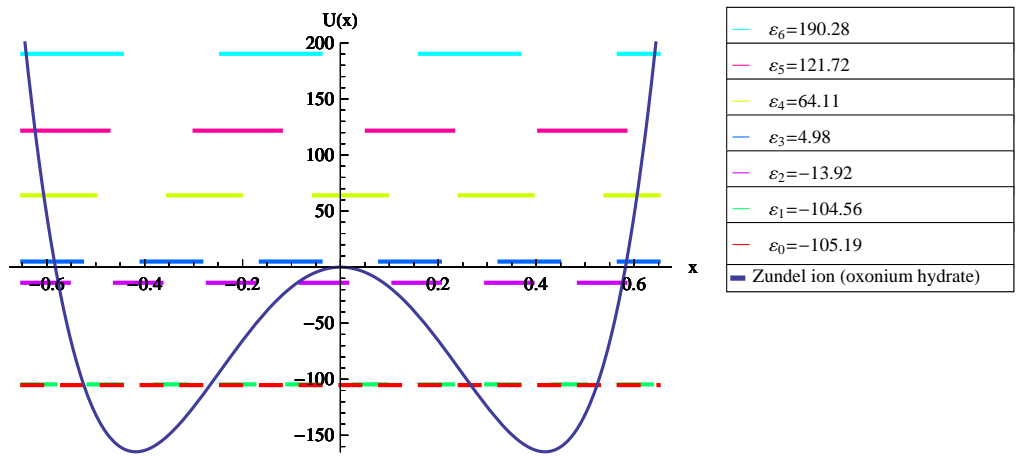}
\end{center}
\caption{The symmetric trigonometric double-well potential at the values of the parameters $m=65$, $p=77.83$. The parameters are chosen to describe the potential and the energy levels for the hydrogen bond in the Zundel ion (oxonium hydrate) for $R_{OO}=2.786\ \AA$
(experimental data are taken from \cite{Yu16}). The increasing energy levels are respectively depicted by the dashes of increasing length.} \label{Fig.5}
\end{figure}
For obtaining the parameters of the symmetric trigonometric DWP ($a=0$) from the data for the Zundel ion presented in \cite{Yu16} we transform the parameters of trigonometric DWP $\{m,p\}$ into the barrier height $B=-U\left(x_{min}\right)$ and width $D=x_{min}^{(1)}-x_{min}^{(2)}$
\[
p=\frac{\sqrt {B}}{1-\left[\cos\left(D/2\right)\right]^2} \ \ \ \ \ \ \ \ \ \ \ \ \ \ \ \ \ \ \ \ \ \ \ m^2-\frac{1}{4}=\frac{B\left[\cos\left(D/2\right)\right]^4}{\left\{1-\left[\cos\left(D/2\right)\right]^2\right\}^2}
\]
Inversely we obtain
\[
D(m,p)=2\arccos\left(\frac{m^2-\frac{1}{4}}{p^2}\right)^{1/4}\ \ \ \ \ \ \ \ \ \ \ B(m,p)=\left(\sqrt {m^2-\frac{1}{4}}-p\right)^2
\]
From Fig.1 of \cite{Yu16} we obtain that for $R_{OO}=2.786\ \AA$ ($2L=2.786\ \AA$) the dimensional distance between the minima of DWP is $X_{min}^{(1)}-X_{min}^{(2)}\approx 0.74\ \AA$. The dimensional barrier height is $V\left(X_{max}\right)-V\left(X_{min}\right)\approx 3100\ {\rm cm^{-1}}$. We obtain that $B\approx 163.02$ and $D\approx 0.83$. From here we have $p\approx 77.83$ and $m=65$. The results are presented in Fig.5.
As a whole the system of energy levels agrees satisfactorily well with that in Fig2. of \cite{Jan72} for the case $R_{OO}=2.8\ \AA$.\\

Acknowledgements. The author is grateful to Dr. Yu.F. Zuev for helpful discussions. The work was partly supported from the government assignment for FRC Kazan Scientific Center of RAS (Project N 0217-2018-0009).

\end{document}